\documentclass[epj]{svjour}
\usepackage{xcolor}
\usepackage{graphicx}
\usepackage{multirow}
\usepackage{eqnarray}
\usepackage{amsmath}
\usepackage{textcomp}
\usepackage[switch]{lineno}

\DeclareGraphicsExtensions{.png,.pdf}
\begin{document}
\title{Ratio of cross-sections of kaons to pions produced in $pp$
  collisions as a function of $\sqrt{s}$}
\author{G.I.~Lykasov, A.I.~Malakhov, A.A.~Zaitsev} 
\institute{Joint Institute for Nuclear Research, Joliot Curie 6, 141980 Dubna, Russian Federation}
\date{Received: date / Revised version: date}
\abstract{A calculation of the inclusive
  spectra of pions
  and kaons produced in
  $pp$ collisions as functions of their 
transverse momentum $p_T$ at mid-rapidity is presented within the
self-similarity approach. 
A satisfactory description of the data within a wide range
of initial energies is presented. We focus mainly on the ratio
of cross-sections of $K^\pm$ to $\pi^\pm$ mesons produced in $pp$
collisions as a function of $\sqrt{s}$.
A fast rise of this ratio, when the initial energy
increases starting from the kaon production threshold up to
$\sqrt{s}\simeq$ 20-30 GeV, is revealed together with its very slow
increase up to LHC energies. The energy dependence of this
ratio is due to the conservation laws of four-momenta and
quantum numbers of the initial and produced hadrons. 
The more or less satisfactory agreement of these ratios with 
NA61/SHINE, RHIC and LHC data is demonstrated. 
}

\PACS{
     {25.40.Fq}{Inelastic proton scattering}   \and
     {13.60.Le}{Meson production }
     }   
\maketitle
\section{Introduction}
\label{intro}

The study of strange hadron production in heavy-ion collisions
compared to the similar production of pions
attracts the attention of theorists and experimenters. This issue
became very important
and intriguing after the observation of a fast rise 
and sharp peak in the ratio of $K^+$ mesons to $\pi^+$ mesons produced in
central $Pb+Pb$ and $Au+Au$ collisions
at mid-rapidity, when the initial energy $\sqrt{s_{NN}}$ per nucleon
grows from the threshold of $K^+$ meson production up to 20-30 GeV
\cite{NA49:2002,NA49:2008} and then falls down at $\sqrt{s_{NN}} >$ 30 GeV. 
However, the energy dependence 
of the $K^-/\pi^-$ ratio in central $Pb+Pb$ and $Au+Au$ collisions is different,
there is no peak at any $\sqrt{s}$, see \cite{NA61/SHINE:2020} and references 
therein.  

The $K^+/\pi^+$ ratio in $pp$ collision at mid-rapidity in contrast to the
similar ratio observed in $AA$ collisions has a fast rise when the initial
energy grows up to $\sqrt{s}\simeq$ 20-30 GeV and then it increases slowly 
with $\sqrt{s}$. The $K^-/\pi^-$ ratio in $pp$
collisions at mid-rapidity has a similar energy dependence,   
see \cite{NA61/SHINE:2020} and references therein.                          

In this paper we analyze the production of kaons and pions in $pp$ collision at
mid-rapidity and focus on ratios between their cross-sections as functions of
the initial energy. Our analysis is based on the similarity of inclusive spectra 
of particles produced in hadron-hadron collisions, suggested in pioneering papers
\cite{Fermi:1950,Pomeran:1951,Landau:1953,Hagedorn:1965}, and on
the conservation laws
of four-momenta and quantum numbers \cite{c3,c4,4,5}.
Actually, we continue to apply the approach developed in recent papers 
\cite{ALM:2015,LM:2018,ML:2020}, where $p_T$-spectra of pions
produced in $pp,AA$ collisions at mid-rapidity and within a
wide range of initial energies were analyzed.

Let us note, that the $p_T$-spectra of hadrons produced in
nuclon-ncleon and nucleus-nucleus collisions were analyzed within the
statistical bootstrap model (SBM)
\cite{Hagedorn:1965,Hagedorn:1973,Hagedorn:1983}. The $K/\pi$ ratios
were calculated in \cite{Hagedorn:1983} as functions of the mean charged
multiplicity emitted by an average cluster or as functions of the average
transverse momentum. However, in our paper only the energy dependence of these 
ratios is analyzed, therefore we cannot compare our results with results
obtained within SBM.

In \cite{c4,5} the similarity was demonstrated at zero 
  rapidity $y=$0 of these spectra as functions of the similarity parameter $\Pi$ dependent
on the initial energy $\sqrt{s}$ in the c.m.s of the colliding
particles and the transverse masses $m_{hT}$ of the produced
hadrons. A simple form of inclusive spectra was used in \cite{c4,4,5} 
to describe satisfactorily the data at low values of $m_{hT}$. Further
development of this approach was presented in our papers \cite{ALM:2015,LM:2018,ML:2020}, 
where the description of $m_{hT}$-spectra was extended to larger values of
transverse masses and initial energies up to a few TeV including
contributions of both quarks and gluons to these spectra.  The relationship between $\Pi$ and the
Mandelstam variables $s,t$ was obtained in \cite{ALM:2015,LM:2018}. Moreover, it has been
shown that $\Pi$ cannot be presented in the factorization form as a common function
of $\sqrt{s}$ and $m_{hT}$. The breakdown of this factorization occurs at not
large initial energies $\sqrt{s}<$ 10 GeV. It is restored at larger $\sqrt{s}$ 
\cite{ALM:2015,LM:2018}. In fact, this is an advantage of the approach based
on the kinematics of four-momentum velocities considered in \cite{c3,c4,4,5}, where the
parameter $\Pi$ was obtained using the conservation laws of four-momenta
and quantum numbers of initial and produced particles, and the minimization
principle. At zero rapidity $y$=0 the form for $\Pi$ was obtained 
analytically \cite{5}.

In this paper we will give a brief review of the main properties of the similarity
approach mentioned above and then present in detail our calculations of transverse
momentum spectra of pions and kaons produced in $pp$ collision within a wide range
  of initial energies $\sqrt{s}$. The main focus of our paper is the theoretical
analysis of ratios of cross-sections of $K^\pm$ to $\pi^\pm$ mesons produced in these
collisions as functions of $\sqrt{s}$ and their comparison with all the world data.    
  
\section{Main properties of the self-similarity approach.}
\label{sec:1}
The inclusive production of hadron $1$ in the interaction of nucleus $A$ with nucleus $B$     
\begin{equation}
A  +  B  \rightarrow  1  + \ldots ,                                           
\label{eq:n1}                                                                
\end{equation}                                              
is satisfied by the conservation law of four-momenta in the following form \cite{4,5}
\begin{equation}
{(N_AP_A + N_BP_{B} - p_1)}^2 = 
{(N_Am_0 + N_B m_0 + M)}^2 ,
\label{eq:n2}
\end{equation}
where $N_A$ and $N_B$ are the fractions of the four-momentum transmitted by
nucleus $A$ and nucleus $B$, their forms are presented in \cite{5,LM:2018} ; 
$P_A$ , $P_B$ , $p_1$ are the four-momenta of nuclei  
$A$ and $B$ and particle $1$,   respectively; $m_0$ is the mass of the nucleon; $M$ is
the mass of the 
particle providing for conservation of the baryon 
number, strangeness, and other quantum numbers.

Let us note, that Eq.~(\ref{eq:n2}) has been introduced in \cite{4} for the
cumulative hadron production in nucleus-nucleus collisions, when the
  kinematics of nucleon-nucleon collisions prohibits hadron production.
In the general case the quantity $M$ should be replaced by the four-momentum 
of the recoil particles. In our case of proton-proton collisions
Eq.~(\ref{eq:n2}) can be valid at initial energies close to the threshold 
production of hadrons. We need this equation to find the minimal value of $M$,
which provides for the conservation of quantum numbers.   
For $\pi$-mesons $m_1 = m_\pi$  and $M = $0.
For anti nuclei $M=m_1$ and for $K^-$-mesons  $M = m_1 = m_K$,
$m_K$ is the mass of the $K$-meson.
For nuclear fragments $M = - m_1$.
For $K^+$-mesons $m_1 = m_K$ and $M = m_\Lambda  - m_K$,
$m_\Lambda$ is the mass of the $\Lambda$-baryon.
Let us note that the isospin effects of the produced hadrons and other nuclear effects
are out of this approach. Therefore, it is assumed that within the self-similarity
approach there is no big
difference between the inclusive spectra of $\pi^+$ and $\pi^-$ mesons produced in $pp$ and $AA$
collisions. However, there is a difference between similar spectra of $K^+$ and $K^-$ mesons, because 
the values of $M$ are different. This is due to the conservation law of strangeness.

 In \cite{4,5} the parameter of self-similarity is introduced in the following 
form
\begin{equation} 
\Pi=\min \left[\frac{1}{2} \left[ (u_A N_A + u_B N_B)^2\right]\right]^{1/2}  ,
\label{eq:n3} 
\end{equation}                                            
where $u_A$ and $u_B$ are the four-velocities of nuclei $A$ and $B$. 
In our case of $pp$ collision $N_A=N_B=N$ is the fraction of the four-momentum
transmitted by   
one proton to another to produce hadron 1. The minimization over 
N presented in Eq.~(\ref{eq:n3}) allows us to find the parameter $\Pi$. This
parameter introduced in \cite{4} was obtained in \cite{5} for nucleus-nucleus
collisions in the mid-rapidity region, however, it can
also be applied successfully for the analysis of 
pion production in $pp$ collisions, as it was shown in
\cite{ALM:2015,LM:2018,ML:2020}. 
Therefore, we continue to apply this method for the analysis of pion
 and kaon production in $pp$ collisions within a wide range of
initial energies.

Then, the inclusive spectrum of particle 1 produced in the $AA$ collision
can be presented as a general universal function dependent on the 
self-similarity parameter $\Pi$ 
\begin{equation}
E d^3 \sigma/dp^3~=~A_A^{\alpha(N_A)}\cdot A_B^{\alpha(N_{B})}\cdot F(\Pi)
\label{eq:n4} 
\end{equation}
where $\alpha(N_A)=1/3 + N_A/3$, $\alpha(N_B)=1/3 + N_B/3$ 
and function $F(\Pi)$ has the following form \cite{ML:2020}
\begin{eqnarray}
F(\Pi)=\bigg[ A_q \mbox{exp}\Big(-\frac{\Pi}{C_q}\Big) + \\
\nonumber
A_g\sqrt{p_T}\phi_1(s) \mbox{exp}\Big(-\frac{\Pi}{C_g}\Big)\bigg] \sigma_{tot}
\label{def:F} 
\end{eqnarray}
where 
\begin{eqnarray}
\Pi(s,m_{1T},y)~=~\left\{\frac{m_{1T}}{2m_0\delta_h}+
\frac{M}{\sqrt{s}\delta_h}\right\}\mbox{cosh}(y)G ,
\label{eq:n10} \\
\nonumber
G = \left\{1+\sqrt{1+\frac{M^2-m_1^2}
{(m_{1T}+2Mm_0/\sqrt{s})^2\mbox{cosh}^2(y)}\delta_h}\right\}~.
\end{eqnarray}
Here 
$\phi_1(s)~=~1-\sigma_{nd}(s)/\sigma_{tot}(s)$, see\cite{LM:2018,ML:2020},\\
$\delta_h=\left(1 - \frac{s_{th}^h}{s} \right)$;
$s_{th}^{\pi}\simeq 4m_0^2$; 
$s_{th}^{K^+}=\left(m_0 + m_K + m_\Lambda \right)^2$;
$s_{th}^{K^-}=(2m_0+2m_K)^2$;
$M = m_\Lambda - m_0; m_\Lambda = $ 1.115 GeV; 
$m_k = $ 0.494 GeV; $s_0 = $ 1 GeV; $m_0 = 0.938$ GeV;
$p_{1T}$ and $m_{1T}$ are the transverse momentum and transverse mass
of the produced hadron $1$; 
$\sigma_{nd} = (\sigma_{tot} - \sigma_{el} - \sigma_{SD})$ is the 
non difrractive cross-section;
$\sigma_{tot},\sigma_{SD}$ and $\sigma_{el}$ are the total
cross-section, the single difrractive cross section and the elastic
cross-section of $pp$ collisions, respectively. They were taken from
\cite{sigma:2013} and \cite{sigm_el:2017} and, together with
 parameters $A_q, C_q$ and  $A_g, C_g$, they are presented in the Appendix.

\section{Transverse momentum spectra and integrated pion and kaon 
cross-sections.}

For $pp\rightarrow h + X$ inclusive processes the relativistic invariant
differential cross-section at small but non-zero rapidity $y$ 
has the following form
\begin{eqnarray}
\rho_h(p_{hT},y)~\equiv E_h\frac{d^3 \sigma_{NN}}{d^3p_1}~=
\frac{1}{\pi}\frac{d\sigma}{dp_{1T}^2dy}~= \\
\nonumber
\frac{1}{\pi}\frac{d\sigma}{dm_{1T}^2dy}=F\left( \Pi(s,m_{1T},y)\right) ,
\label{eq:n11}
\end{eqnarray}
where $F(\Pi(s,m_{1T},y))$ and $\Pi(s,m_{1T},y)$ are given by \\
Eqs.~(5,6). 

The production cross-section of  hadron $h$ integrated over its
transverse momentum $p_{1T}$ or transverse mass $m_{1T}$ at zero rapidity
$y=0$ and $s\geq s_{th}^h$ can be presented in the following form
\begin{eqnarray}
\frac{d\sigma_h}{dy}(s,y=0)
=2\pi\int_{p^{min}_{1T}}^{p^{max}_{1T}}\rho(s,p_{1T},y=0)p_{1T}dp_{1T}= \\
\nonumber
2\pi\sigma_{tot}(s)\int_{p^{min}_{1T}}^{p^{max}_{1T}}\mbox{exp}\left( -\frac{M}{C_q\sqrt{s}\delta_h}G\right)(J_q+J_g)p_{1T}dp_{1T} ,  
\label{eq:crsec}
\end{eqnarray}
where
\begin{eqnarray}
J_q~=~A_q\mbox{exp}\Bigg(-\frac{m_{1T}}{2m_0C_q\delta_h}G \Bigg)~
\label{eq:Jq}
\end{eqnarray}
and
\begin{eqnarray}
J_g~=~A_g\mbox{exp}\Bigg(\frac{M(C_g-C_q)}{C_qC_g\sqrt{s}\delta_h}G\Bigg)\sqrt{p_{1T}}\phi_1(s)\times\\
\nonumber
\mbox{exp}\Bigg(-\frac{m_{1T}}{2m_0C_g\delta_h}G \Bigg)~
\label{eq:Jg}
\end{eqnarray}
Let us analyze qualitatively the energy behaviors of pion and kaon production
cross-sections in $pp$ collisions at zero rapidity 
given by Eqs.~(8-10).  Their $\sqrt{s}$ dependence is determined mainly by the
factor $\mbox{exp}\big(-\frac{M}{C_q\sqrt{s}\delta_h}G\big)$
and factors $\mbox{exp}\big(-\frac{m_{1T}}{2m_0C_q\delta_h}G\big)$,
$\mbox{exp}\big(-\frac{M(C_g-C_q)}{C_qC_g\sqrt{s}\delta_h}G\big)$,
$\mbox{exp}\big(-\frac{m_{1T}}{2m_0C_g\delta_h}G\big)$
entering into functions $J_q$ and
$J_g$. For all hadrons produced at the threshold these factors result in the
zero cross section due to zero of the kinematic factor $\delta_h=1-s_{th}/s$
at $s=s_{th}$. For $\pi^\pm, K^+, K^-$ mesons the threshold
energies $\sqrt{s_{th}}$ are 2.015 GeV, 2.547 GeV and 2.86 GeV, respectively.
At $s>s_{th}$ the production cross-sections of $K^\pm$ mesons show a
fast rise due to $M$ not being equal to zero and to an increase
of $\delta_h=1-s_{th}/s$, then at $s\gg s_{th}$ and
$s\gg M$ the energy dependence of these cross-sections changes only due to the
total cross section $\sigma_{tot}(s)$ and 
$\phi_1(s)~=~1-\sigma_{nd}(s)/\sigma_{tot}(s)$ \cite{LM:2018,ML:2020}
entering into Eq.~(10). The energy dependence of the production
cross-section of pions in $pp$ collision is similar to the kaon production
cross-section, however, the threshold of the pion production 
$s_{th}^\pi$ is less than the kaon one $s_{th}^{K^\pm}$, as it is mentioned
above. Therefore, the ratio of cross-sections, $\sigma_{K^\pm}/\sigma_\pi$\,
exhibits a fast
rise from the kaon threshold when $\sqrt{s}$ grows due to an increase of the
phase space. Then, at $s\gg s_{th}$ this rise is broken and goes to a slow
increase due, mainly, to the factor 
$\mbox{exp}\big(\frac{M(C_g-C_q)}{C_qC_g\sqrt{s}\delta_h}\big)$ presented in Eq.~(10)
because $C_g>C_q$, as it is shown in Table 1 of the Appendix.   

\begin{figure}[hbtp] 
\begin{center}
\includegraphics[width=0.5\textwidth]{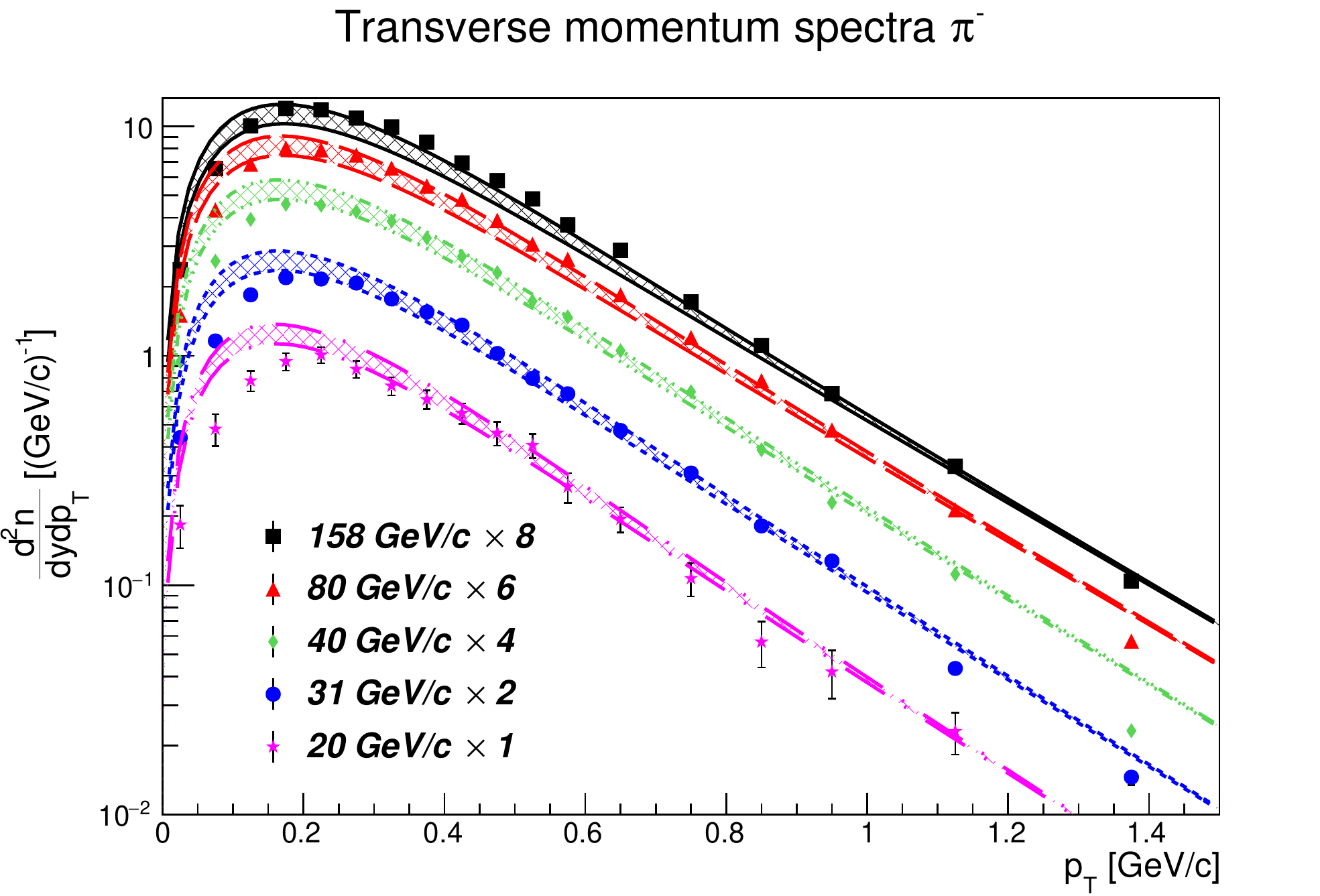}
\end{center}
 \caption{
The $p_T$-spectra of $\pi^-$ mesons produced in $pp$ collision at $P_{in}=$
158 GeV$/$c ($\sqrt{s}$=17.28 GeV),
80 GeV$/$c ($\sqrt{s}$ = 12.34 GeV), 40  GeV$/$c ($\sqrt{s}$ = 8.77 GeV), 31
GeV$/$c ($\sqrt{s}$ = 7.75 GeV),
20  GeV$/$c ($\sqrt{s}$ = 6.27 GeV) at mid-rapidity $y<$ 0.2. The lines are
our calculations, data are taken from 
\cite{NA61:2017}, the bands are due to uncertainties in
parameter $A_q$ presented
in the Appendix.
}  
\label{fig_pimin}
\end{figure}
\begin{figure}[hbtp] 
\begin{center}
\includegraphics[width=0.5\textwidth]{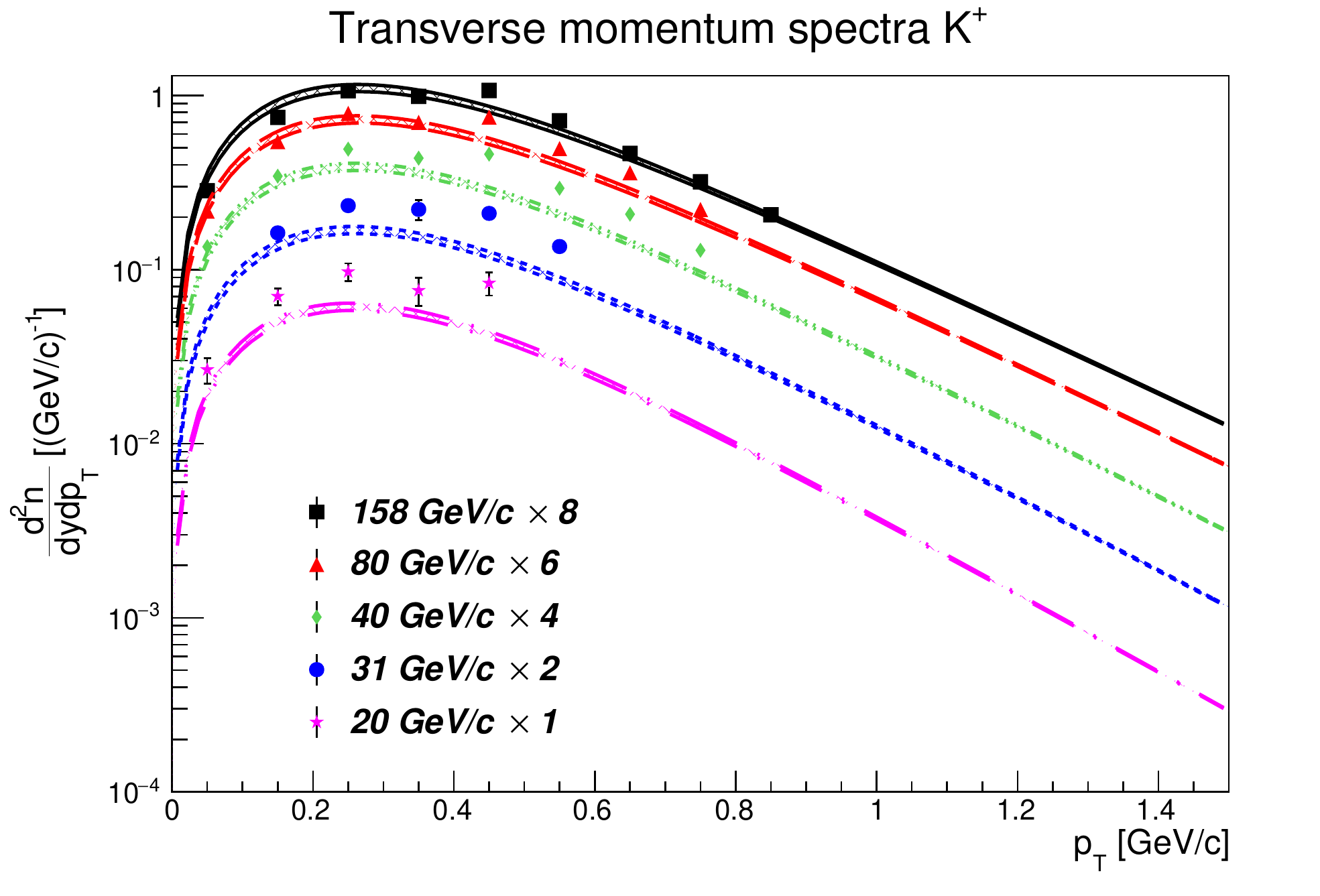}
\end{center}
 \caption{
The $p_T$-spectra of $K^+$ mesons produced in $pp$ collision.
Notations are the same as in Fig.~\ref{fig_pimin}.
}  
\label{fig_Kplus}
\end{figure}
\begin{figure}[hbtp] 
\begin{center}
\includegraphics[width=0.5\textwidth]{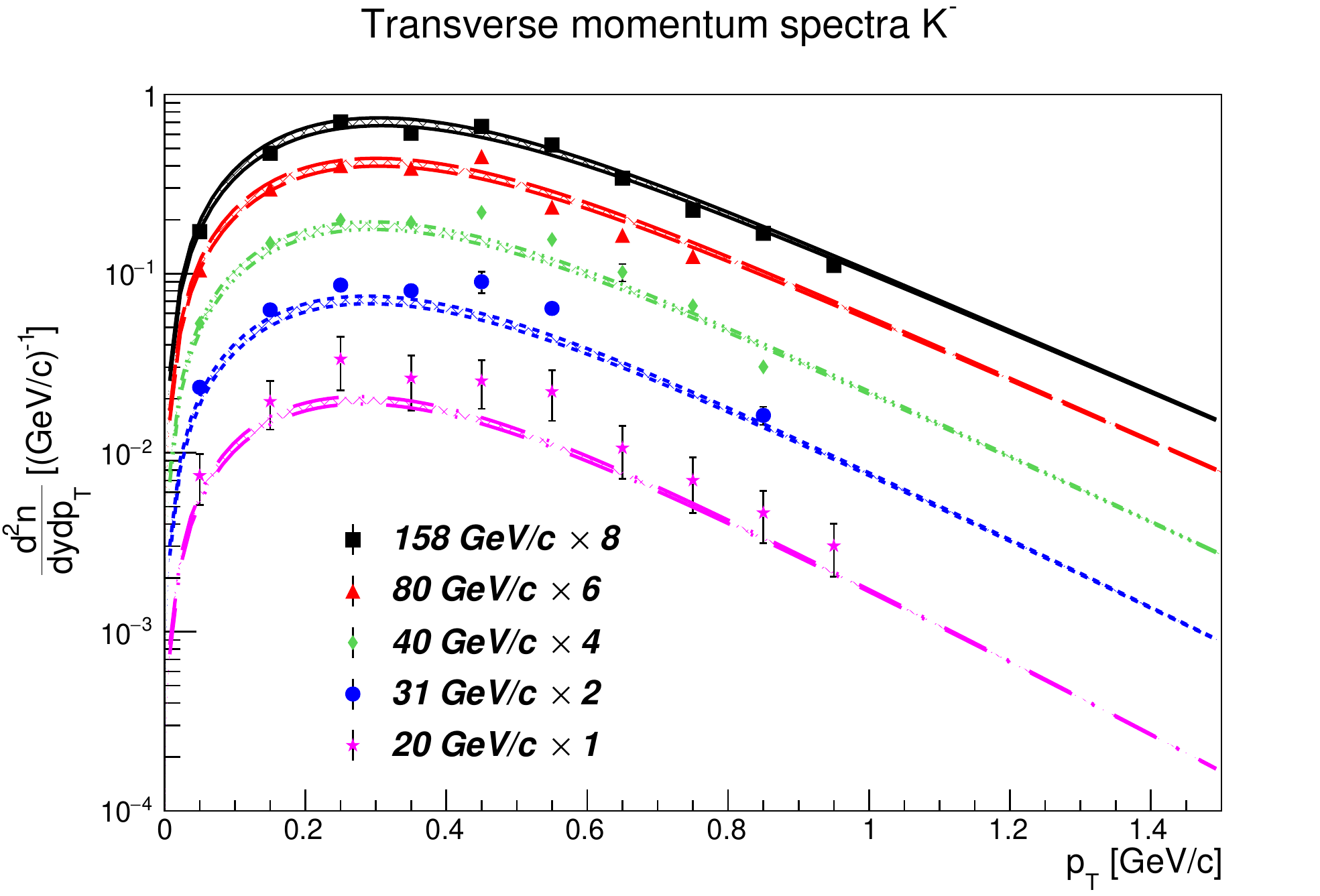}
\end{center}
 \caption{
The $p_T$-spectra of $K^-$ mesons produced in $pp$ collision.
Notations are the same as in Fig.~\ref{fig_pimin}.
}  
\label{fig_Kmin}
\end{figure}
The $p_T$-spectra of $\pi^-$, $K^+$ and $K^-$ mesons are sums of quark and gluon
contributions including uncertainties due to the fit of data are presented in
Figs.~(\ref{fig_pimin}-\ref{fig_Kmin}). 
By fitting NA61/SHINE data on $p_T$-spectra at mid-rapidity the parameters
$C_q,A_g,C_g$ were found to be independent of the initial energy $\sqrt{s}$,
they depend on the kind of mesons produced, $\pi,K^+,K^-$.
However, the parameter $A_q$ varies a little bit at energies from 40 GeV$/$c up to
158 GeV$/$c. The uncertainties in $p_T$-spectra and ratios of yields, $K^+/\pi^+$ and
$K^-/\pi^-$ are due to the uncertainties in the parameter $A_q$.
All these parameters are presented in the Appendix. The similar spectra 
with quark and gluon contributions are also presented in the Appendix.
\section{Ratio of the kaon and pion yields.}

\begin{figure}[hbtp] 
\begin{center}
\includegraphics[width=0.5\textwidth]{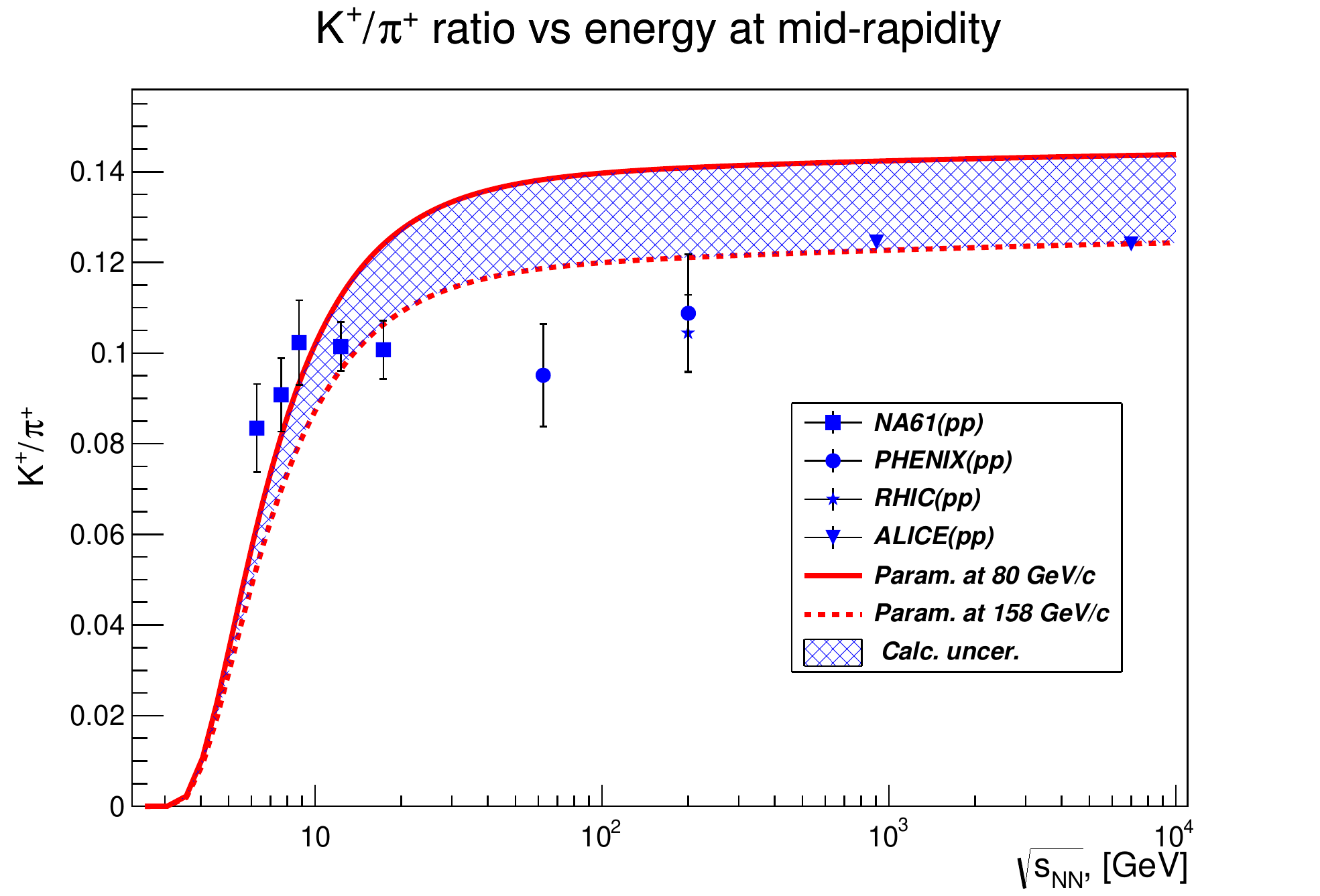}
\end{center}
 \caption{The ratio between yields of $K^+$ and $\pi^+$ mesons produced in $pp$ 
collisions as a function of $\sqrt{s}$.
 }  
\label{fig_ratKplpipl}
\end{figure}

\begin{figure}[hbtp] 
\begin{center}
\includegraphics[width=0.5\textwidth]{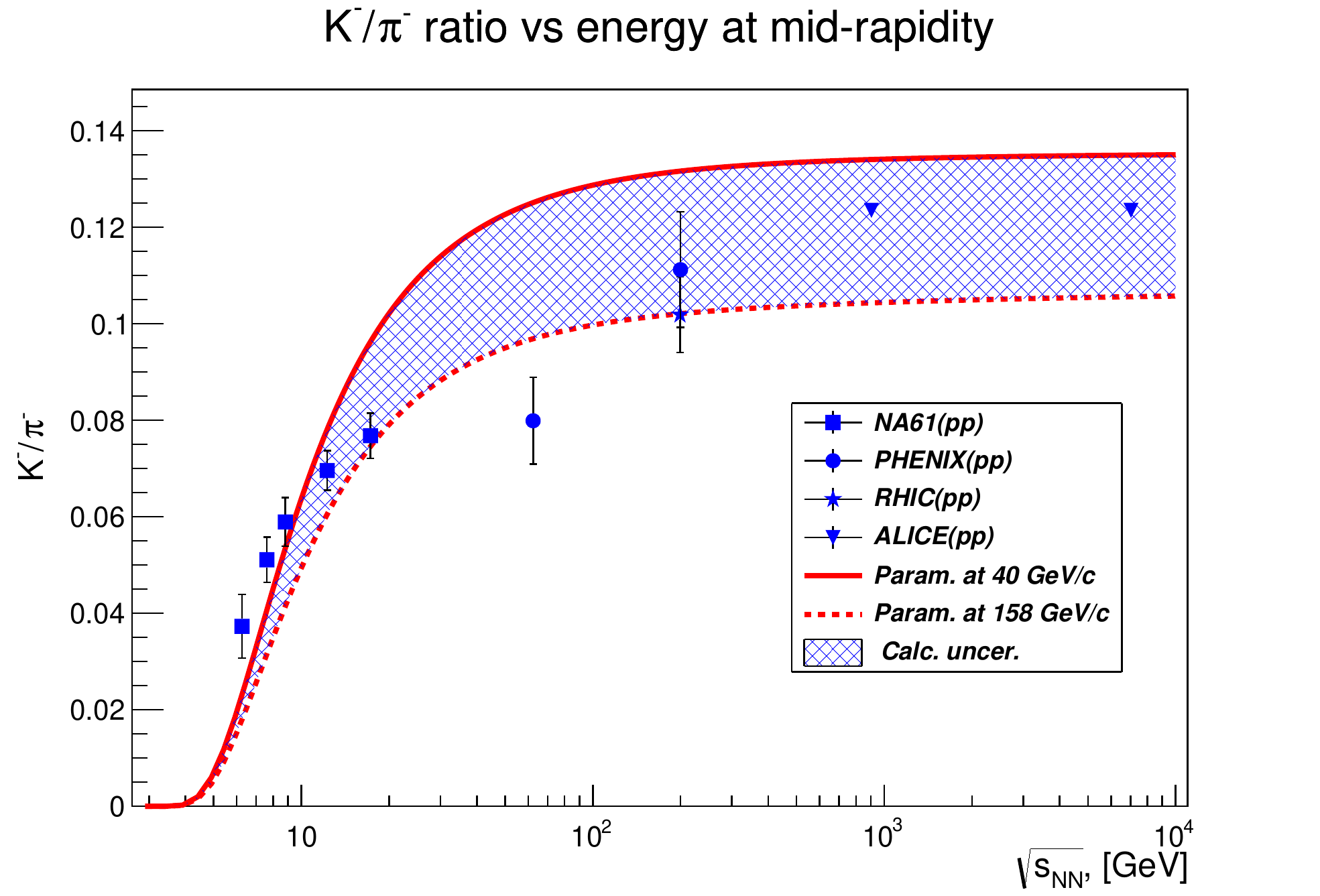}
\end{center}
 \caption{The ratio between yields of $K^-$ and $\pi^-$ mesons produced in $pp$ 
collisions as a function of $\sqrt{s}$.
}  
\label{fig_ratKmin_20}
\end{figure}

In Figs.~(\ref{fig_ratKplpipl},\ref{fig_ratKmin_20}) the respective yield
ratios, $K^+/\pi^+$ and $K^-/\pi^-$, are presented as functions 
of $\sqrt{s}$. 
From these figures one can see their fast rise from the threshold energy of
$K^+$ or $K^-$ production up to $\sqrt{s}=$ 20-30 GeV and
their further slow increase as the energy grows. 

The upper line in Fig.~\ref{fig_ratKplpipl} corresponds to the fit of data for
$\pi^+$ and $K^+$ mesons at $P_{in}=$ 80 GeV$/$c, and the bottom line corresponds to
the similar fit at
$P_{in}=$ 158 GeV$/$c. The upper curve in Fig.~\ref{fig_ratKmin_20} corresponds to the
fit of data for 
$\pi^-$ and $K^-$ mesons at $P_{in}=$ 40 GeV$/$c and the bottom line corresponds to the similar fit at
$P_{in}=$ 158 GeV$/$c.
  
\section{Conclusion}

In this paper we have applied the self-similarity approach of analysis of
hadron production in $pp$ collisions to the 
production of both kaons and pions in $pp$ collisions at mid-rapidity 
$y<0.2$ within a wide range of initial energies.   
The main goal of this paper is to analyze the ratio of kaon yields  to
those of pions, produced in $pp$ collisions, within
this approach as a function of $\sqrt{s}$. 
We have presented a self-consistent satisfactory description
of the data on $p_T$-spectra of pions and kaons in a wide range of initial energies
and not large transverse momenta. The fast rise of 
$K^+/\pi^+$ and $K^-/\pi^-$ yield ratios as functions of $\sqrt{s}$ 
from the threshold energy of $K^+$ or $K^-$ production up to $\sqrt{s}=$ 20-30 GeV 
has been demonstrated as well as their further slow increase with growing energy. 
The energy dependence of these ratios calculated within the
suggested approach results in a satisfactory description
of the data presented by 
NA61/SHINE at $6.27\leq\sqrt{s}\leq 17.28$ GeV. As for the descriptions of data 
on $K^\pm/\pi^{\pm}$ yield ratios measured by PHENIX, STAR and ALICE
Collaborations, they do not contradict data taking into account the uncertainty
of parameter $A_q$ presented in Figs~(\ref{fig_ratKplpipl}-\ref{fig_ratKmin_20}) 
as shade bands.

The physical reason of this energy dependence of the ratio of
kaon yields to those of pions consists in the conservation
laws of four-momenta and of reaction
quantum numbers.  
The conservation law of the four-momenta
of initial and produced particles results in the fast
rise with energy of the kaon and pion production  
cross-sections, when $\sqrt{s}$ grows from the threshold energy.
This is due to the factor $\delta_h=1-s^h_{th}/s$ entering into the
self-similarity function $\Pi(s,m_{1T},y)$ given by Eq.(6). A similar
fast rise of the $K^{\pm}/\pi^\pm$ yield ratios is also due
to the factor $\delta_h$ and to the non-zero  
value of $M$ for $K^+$ and $K^-$ mesons that also enters into
$\Pi(s,m_{1T},y)$. When $\sqrt{s}\gg\sqrt{s_{th}}$ and $\sqrt{s}\gg M$,
the pion and kaon production cross-sections and their ratios
become insensitive to factors $\delta_h$ and $M$, however they are sensitive to the
difference between the quark and gluon contributions to the pion and kaon
spectra as functions of $p_T$ and $\sqrt{s}$. 
That is why the $K^\pm/\pi^\pm$ yield ratios exhibit
two kinds of energy dependence, a fast rise,
when $\sqrt{s_{th}}<\sqrt{s}<$ 20-30 GeV and a slow increase, when 
$\sqrt{s}>$ 20-30 GeV. 

Let us note that no fast rise and sharp peak in the
ratio between the yields of $K^+$ and 
$\pi^+$ mesons produced in central $BeBe$ collisions are
observed in the NA61/SHINE experiment, according to
\cite{NA61:2020}. This ratio is very similar to the same
$K^+/\pi^+$ ratio measured in $pp$ collisions by the
NA61/SHINE Collaboration. Therefore, we guess that the approach, presented in
this paper, applied in the analysis of pion
and kaon production in the collisions of light nuclei can
give results similar to those for $pp$
collisions. At present, this work is in progress.
However, the self-similarity approach can, probably, not
explain the anomalous $K^+/\pi^+$ ratio in heavy-ion collisions
(Pb-Pb, Au-Au) around collision energies of 20-30 GeV, because
it assumes the same forms (slopes) of particle spectra in p+p
and A+A collisions.


\section{Appendix} 
The parameterizations of $\sigma_{tot},\sigma_{SD}$ and $\sigma_{el}$ 
have the following forms \cite{sigma:2013} and \cite{sigm_el:2017} \\
$\sigma_{tot} = (21.7(s/s_0)^{0.0808} + 56.08(s/s_0)^{-0.4525}$) mb;\\ 
$\sigma_{el} = (12.7 - 1.75\mbox{ln}(s/s_0) + 0.14\mbox{ln}^2(s/s_0)$) mb;\\ $\sigma_{SD} = (4.2 +
\mbox{ln}(\sqrt{s/s_0})$) mb.

\begin{figure*}[h]
	\centering
	\includegraphics[width=5.0in]{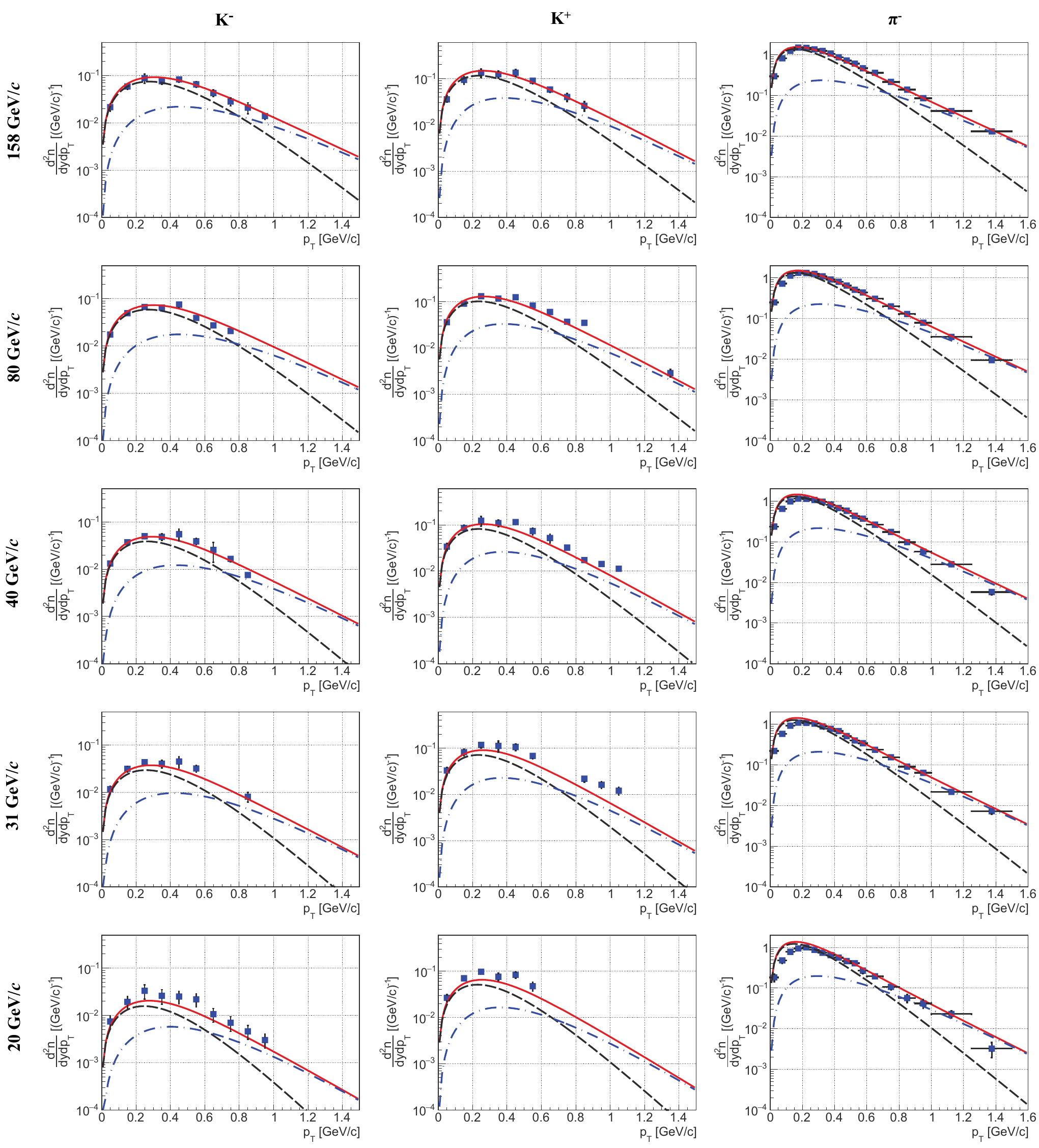}
	\caption{The $p_T$ spectra of $\pi^-$, $K^+$ and $K^-$ mesons produced
          at y $\approx$ 0 in inelastic $pp$ interactions at SPS energies 
        $\sqrt{s}=$ 6.3 - 17.3 GeV or $P_{in}=$ 20-158 GeV$/$c. Data are taken
          from \cite{NA61:2017}.}
	\label{fig_pimin_158}
\end{figure*}

In Fig.~\ref{fig_pimin_158} the $p_T$-spectra of pions and
kaons, produced in $pp$ collisions within the
initial momentum range (20-158) GeV/c, fitted by NA61/SHINE data
are presented. The black dashed line corresponds
to the quark contribution, the blue dash-dotted curve is the gluon
contribution and the red solid line is the sum of quark and nonperturbative gluon
contributions. The parameters $A_q,A_g$ and $C_q,C_g$ were found from
a fit of NA61/SHINE data and are presented
in Table 1.

As it is shown in \cite{ALM:2015,LM:2018,ML:2020}, the form of
inclusive pion spectra versus $p_T$ at mid-rapidity given by Eqs.~(4-6) 
describes satisfactorily data in a wide range of $\sqrt{s}$ at $p_T<$ 2-3
GeV$/$c. Moreover, as it is shown in \cite{BGLP:2012,GLLZ:2013,LLZ:2014} and
\cite{AJLLM:2018}, the contribution of gluons to the pion
spectrum is related to the gluon
distribution at low $Q^2=$ 1-2 (GeV$/$c)$^2$, the use of which results in 
a satisfactory description of data on hard $pp$ processes at
LHC energies and of proton structure functions at low $x$.
Therefore,  we use Eqs.~(4-7) for description of data on pion $p_T$-spectra in
$pp$ collisions, only  
improving the fit of data. 

As for $K^\pm$ production in $pp$ collisions at
not large initial energies
we take into account the additional contribution due to the one Reggeon exchange
diagram, which has $\sqrt{s_{th}/s}$ dependence. It leads to
modification of parameter $A_q$ in the following form $A_q(1+\sqrt{s_{th}/s})$, 
which can be approximated by $A_q\mbox{exp}(\sqrt{s_{th}/s})$. This correction 
vanishes at RHIC and LHC energies, however, it allows 
us to describe data at $\sqrt{s}<$ 10 GeV satisfactorily. 

The parameter $A_q$ for $\pi$ meson
production was found from the fit of NA61 data
\cite{NA61:2017,NA61/SHINE:2020} at initial energies $P_{in}=$ 40-158 GeV$/$c. 
It is very close to the value of $A_q$ obtained in 
\cite{ALM:2015,LM:2018,ML:2020}. For $K^+$ production the value of $A_q$ was
found from the fit of NA61 data at $P_{in}=$ 80 GeV$/$c and $P_{in}=$ 158 GeV$/$c. For
$K^-$ $A_q$ was found from a fit of NA61 data at $P_{in}=$ 40 GeV$/$c
and $P_{in}=$ 158 GeV$/$c. 
Other parameters $A_g, C_q$ and $C_g$ were fixed 
from a fit of the NA61 data at $P_{in}=$ 80 GeV$/$c and they do not depend on
other initial energies.   

\begin{table*}[h]
\centering
\caption{Table of parameters found from the fit of NA61/SHINE data \cite{NA61:2017}.}
\begin{tabular}{ | c | c | c | c | c | c | c | c |}
\hline
$pp \to h X$ & \multicolumn{3}{c|}{$\pi^-$} & \multicolumn{2}{c|}{$K^{+}$} & 
\multicolumn{2}{c|}{$K^{-}$}   \\ \hline
$\sqrt{s_{pp}}$, GeV & 17.3 & 12.3 & 8.8 & 17.3 & 12.3 & 17.3 & 8.8 \\ \hline
$\textit{P}$, GeV/$c$ & 158 & 80 & 40 & 158 & 80 & 158 & 40 \\ \hline
A$_{q}$ & 3.361 & 3.063 & 2.688 & 0.9925 & 1.152 & 1.951 & 2.219 \\ \hline
C$_{q}$ & \multicolumn{3}{c|}{0.147} & \multicolumn{2}{c|}{0.148} & \multicolumn{2}{c|}{0.148} \\ \hline
A$_{g}$
& \multicolumn{3}{c|}{1.788} & \multicolumn{2}{c|}{0.7726} & \multicolumn{2}{c|}{0.629} \\ \hline
C$_{g}$ & \multicolumn{3}{c|}{0.22} & \multicolumn{2}{c|}{0.2066} & \multicolumn{2}{c|}{0.2271} \\ 
\hline
\end{tabular}
\label{table 1}
\end{table*}

{\bf Acknowledgements.}

\begin{sloppypar} 
We are very grateful to K.A. Bugaev, M. Gumberidze, M. Gazdzicki, R. Holzmann,
S. Pulawski, G.Pontecorvo for extremely helpful discussions.   
\end{sloppypar}

\end{document}